\documentclass[12pt,b5paper]{article}
\usepackage{amsmath,amssymb,amsfonts,amsthm,amsopn,graphicx,appendix}
\usepackage{tikz} 
\usetikzlibrary{arrows}
\allowdisplaybreaks[3]
\numberwithin{equation}{section} 
\newtheorem{theorem}{Theorem}[section]

\newtheorem{statement}[theorem]{Statement}

\newcommand{\ds}{\displaystyle}
\def\EXP{\textrm{{\large e}}}
\newcommand{\ii}{\mathsf{i}}
\newcommand{\pt}{\boldsymbol{n}}
\newcommand{\one}{\boldsymbol{e}_1}
\newcommand{\two}{\boldsymbol{e}_2}
\newcommand{\three}{\boldsymbol{e}_3}

\renewcommand{\author}[1]{\large\rm #1\\ \bigskip}
\newcommand{\address}[1]{{\normalsize\it #1\\}\bigskip}
\renewcommand{\title}[1]{\bigskip\bigskip\Large\bf #1\bigskip\bigskip\\}
\usepackage[body={15.5cm,22cm}]{geometry}
\newcommand{\uop}{\boldsymbol{u}}
\newcommand{\vop}{\boldsymbol{v}}

\def\PAC{blue!60!black}

\def\GR{red!70!blue}

\newcounter{app}
\newcounter{sapp}[app]

\begin{document}
\vglue 2cm

\begin{center}

\title{On Pentagon Equation, Tetrahedron Equation, Evolution and Integrals of Motion.}
\author{Sergey M.~Sergeev.}

\vspace{.5cm}

\address{Department of Theoretical Physics,
         Research School of Physics and Engineering,\\
    Australian National University, Canberra, ACT 0200, Australia\\
    and\\
   Faculty of Science and Technology, \\
   University of Canberra, Bruce ACT 2617, Australia }

%\date{}

\end{center}
%\maketitle
%%%%%%%%%%%%%%%%%%%%%%%%%%%%%%%%
\begin{abstract}
There is a sub-class of the solutions to Quantum Tetrahedron Equation related to the algebraical Pentagon Equation. The Quantum Tetrahedron Equation defines an evolution operator in wholly discrete three dimensional space-time. In this paper we establish the Liouville integrability of one particular quantum evolution model/classical integrable model on cubic lattice. The key feature of the model is that it has two independent quantum/classical spectral curves. In particular, on the classical level its Hamiltonian equations of motion decouple into two independent Hirota equations.
\end{abstract}

%%%%%%%%%%%%%%%%%%%%%%%%%%%%%%%%

%%%%%%%
%%%%%%%%%%%%%%%%%%%%%%%%%%%

\section{Introduction}

We discuss in this paper some specific three-dimensional exactly integrable model related to the algebraic Pentagon equation.

In particular, we consider the solution of the Tetrahedron equation 
\begin{equation}\label{TE}
R_{123}^{} R_{145}^{} R_{246}^{} R_{356}^{} \;=\; R_{356}^{} R_{246}^{} R_{145}^{} R_{123}^{}
\end{equation}
of the form
\begin{equation}\label{R}
R_{123}^{} \;=\; S_{13}^{} P_{23}^{} S_{13}^{-1}\;,
\end{equation}
where $P_{23}$ is the permutation operator, while $S_{13}$ is the subject of the algebraic Pentagon equation
\begin{equation}\label{PE}
S_{12}^{} S_{23}^{} \;=\; S_{23}^{} S_{13}^{} S_{12}^{}\;,
\end{equation}
and an auxiliary $10$-term relation (see \cite{KS1} for details),
\begin{equation}\label{10}
S_{12}^{} S_{13}^{-1} S_{14}^{} S_{24}^{-1} S_{34}^{} \;=\;
S_{24}^{-1} S_{34}^{} S_{14}^{-1} S_{12}^{} S_{13}^{-1}\;.
\end{equation}

The $R$-matrix of the Tetrahedron Equation can be used for construction of the evolution operator, its kernel is given by
\begin{equation}\label{U}
U_{\{\;\alpha,\;\beta,\;\gamma\}}^{\{\alpha',\beta',\gamma'\}}\;=\;
\prod_{i,j} R_{\;\;\;\;\alpha_{i,j},\;\;\beta_{i,j},\;\;\;\;\;\;\gamma_{i,j}}^{\alpha_{i+1,j}',\;\;\beta_{i,j}',\;\;\gamma_{i,j+1}'}\;.
\end{equation}
The periodical boundary conditions imply $i,j\in\mathbb{Z}_N$. Evolution operator commutes with a layer-to-layer transfer-matrix constructed with the help of $R$-matrices, usually this justifies the integrability in the case when $R$-matrix is endowed by a set of spectral parameters.

In this paper we consider the case of the constant Pentagon and Tetrahedron Equations related to the Weyl algebra of observables and quantum dilogarithm. We establish the complete integrability of the model by producing explicitly the complete set of the integrals of motion. The remarkable  feature of the model is that the complete set is produced by two independent determinants (generating functions) rather then by one. On the classical level this means that the model has two independent spectral curves.

\section{Linear problems and Tetrahedral map}

We commence with the formulation of an auxiliary linear problem(s) intertwined by $R$-matrix (\ref{R}) and allowing to derive the quantum spectral curves for the quantum evolution operator (\ref{U}).

A remarkable feature of $R$-matrix (\ref{R}) is that there are two completely different approaches to its linear problems. One is based on the linear problem providing the algebraic Pentagon equation (\ref{PE}). This approach is rather geometrical, but it does not provide a way to produce a set of integrals of motion. We will give details in the appendix.

The auxiliary linear problem used in this section can be formulated as follows. The picture below gives a combinatoric of a Tetrahedral map:
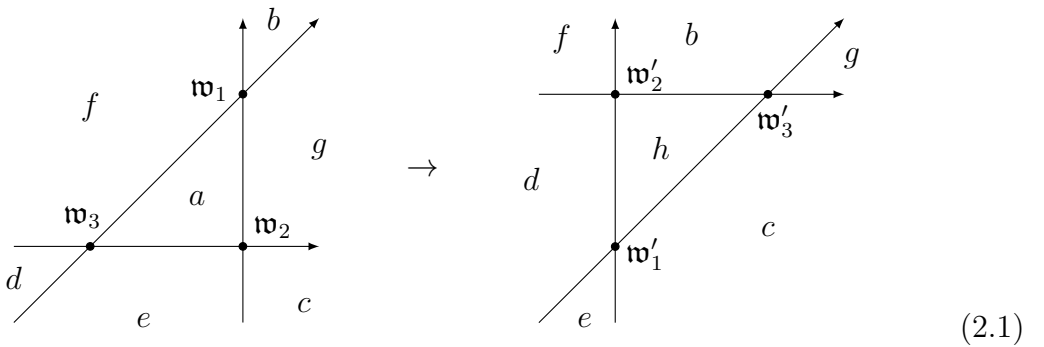
\begin{equation}\label{LYBE}
\begin{tikzpicture}
\draw [-latex, thin] (-3,0) -- (1,0);
\draw [-latex, thin] (-3,-1) -- (1,3);
\draw [-latex, thin] (0,-1) -- (0,3);
\node [above] at (-0.6,0.4) {$a$};
\node [above] at (-2,1.5) {$f$};
\node [above] at (-3,-0.7) {$d$};
\node [above] at (0.4,2.7) {$b$};
\node [above] at (1,1) {$g$};
\node [above] at (0.8,-1) {$c$};
\node [above] at (-1.3,-1.2) {$e$};
\draw [fill] (0,2) circle [radius=0.05];
\draw [fill] (0,0) circle [radius=0.05];
\draw [fill] (-2,0) circle [radius=0.05];
\node [left] at (-0.05,2.05) {$\mathfrak{w}_1$};
\node [above] at (-2.1,0.1) {$\mathfrak{w}_3$};
\node [right] at (0.,0.25) {$\mathfrak{w}_2$};
\node [right] at (2,1) {$\to$};
\end{tikzpicture}
\qquad
\begin{tikzpicture}
\draw [-latex, thin] (-1,0) -- (3,0);
\draw [-latex, thin] (-1,-3) -- (3,1);
\draw [-latex, thin] (0,-3) -- (0,1);
\node [above] at (0.6,-1) {$h$};
\node [above] at (-0.7,0.4) {$f$};
\node [above] at (-1.1,-1.4) {$d$};
\node [above] at (1,0.5) {$b$};
\node [above] at (3.1,0.2) {$g$};
\node [above] at (2,-2) {$c$};
\node [above] at (-0.4,-3.2) {$e$};
\draw [fill] (0,-2) circle [radius=0.05];
\draw [fill] (0,0) circle [radius=0.05];
\draw [fill] (2,0) circle [radius=0.05];
\node [right] at (0.,-2.1) {$\mathfrak{w}_1'$};
\node [below] at (2.1,-0.) {$\mathfrak{w}_3'$};
\node [right] at (0.,0.25) {$\mathfrak{w}_2'$};
\end{tikzpicture}
\end{equation}
On this picture $a,b,\cdots,h$ denote the cells of two two-dimensional kagome lattices\footnote{Kagome lattice is a lattice formed by three sets of parallel lines. Kagome lattice also can be seen as a dissection of a cubic lattice by an auxiliary plane in general position. R. J. Baxter suggested the term ``kagome lattice'' inspired by the Japanese handcraft} . The vertices stand for local Weyl algebras. The left and right graphs of (\ref{LYBE}) present two systems of linear equations:
\begin{equation}\label{LP1}
\left\{ \begin{array}{l}
\ds \ell_1^{} \; \stackrel{def}{=}\; \phi_f^{} - \vop_1^{} \phi_a^{} + \uop_1^{} \phi_b^{}\;=\;0\\
\\
\ds \ell_2^{} \; \stackrel{def}{=}\; \phi_a^{} - \vop_2^{} \phi_e^{} + \uop_2^{} \phi_g^{} \;=\; 0\\
\\
\ds \ell_3^{} \; \stackrel{def}{=}\; \phi_f^{} - \vop_3^{} \phi_d^{} + \uop_3^{} \phi_a^{} \;=\; 0
\end{array}
\right. 
\quad \to
\quad
\left\{ \begin{array}{l}
\ds \ell_1^{\prime} \; \stackrel{def}{=}\; \phi_d^{} - \vop_1' \phi_e^{} + \uop_1' \phi_h^{} \;=\; 0\\
\\
\ds \ell_2^{\prime} \; \stackrel{def}{=}\; \phi_f^{} - \vop_2' \phi_d^{} + \uop_2' \phi_b^{} \;=\; 0\\
\\
\ds \ell_3^{\prime} \; \stackrel{def}{=}\; \phi_b^{} - \vop_3' \phi_h^{} + \uop_3' \phi_g^{} \;=\; 0
\end{array}
\right.
\end{equation}
Here $\uop,\vop$ stand for a local Weyl algebra,
\begin{equation}
\mathfrak{w}_j\;=\;( \uop_j,\; \vop_j\; : \; \uop_j\vop_j\;=\;q^2\vop_j\uop_j)
\end{equation}
(the term ``local'' means that the Weyl elements assigned to different vertices on the same graph commute).  Auxiliary elements $\phi_a,\cdots,\phi_h$ belong to a formal right module of all Weyl algebras.

The intertwining means the linear equivalence of the left and right systems of (\ref{LP1}). In particular, $\vop_1^{-1}\ell_1^{}+\uop_3^{-1}\ell_3^{}$ and $\ell_2'$ have the same auxiliary variables and therefore they must be proportional,
\begin{equation}
(\vop_1^{-1}+\uop_3^{-1})^{-1} (\vop_1^{-1}\ell_1^{}+\uop_3^{-1}\ell_3^{}) \;\equiv\; \ell_2'\;.
\end{equation}
The equivalence of two linear systems (\ref{LP1}) provides
\begin{equation}
\begin{array}{lll}
\ds \uop_1'\;=\;\vop_3^{-1} F\;, & \ds \uop_2'\;=\;\uop_3^{} (\vop_1^{}+\uop_3^{})^{-1}\uop_1^{}\;, & \ds \uop_3^{} \;=\; \uop_1^{-1}(\vop_1^{}+\uop_3^{}) \uop_2^{}\;,\\
\\
\ds \vop_1'\;=\;\vop_3^{-1}(\vop_1^{}+\uop_3^{}) \vop_2^{}\;, & \ds \vop_2'\;=\;\vop_1^{}(\vop_1^{}+\uop_3^{})^{-1}\vop_3^{}\;, & \ds \vop_3'\;=\;\uop_1^{-1} F\;,
\end{array}
\end{equation}
where $F$ remains undefined.

There are several possible choices of $F$ providing the integrability \cite{S99}. In this paper we fix $F$ by an extra requirement: the same $\mathfrak{w}_j^{},\mathfrak{w}_j'$ must satisfy the second independent linear problem:
\begin{equation}\label{LP2}
\left\{ \begin{array}{l}
\ds \tilde{\ell}_1^{} \; \stackrel{def}{=}\; \tilde{\phi}_g^{} - \vop_1^{} \tilde{\phi}_b^{} + \uop_1^{} \tilde{\phi}_a^{}\;=\;0\\
\\
\ds \tilde{\ell}_2^{} \; \stackrel{def}{=}\; \tilde{\phi}_c^{} - \vop_2^{} \tilde{\phi}_g^{} + \uop_2^{} \tilde{\phi}_e^{} \;=\; 0\\
\\
\ds \tilde{\ell}_3^{} \; \stackrel{def}{=}\; \tilde{\phi}_e^{} - \vop_3^{} \tilde{\phi}_a^{} + \uop_3^{} \tilde{\phi}_d^{} \;=\; 0
\end{array}
\right. 
\quad \to
\quad
\left\{ \begin{array}{l}
\ds \tilde{\ell}_1^{\prime} \; \stackrel{def}{=}\; \tilde{\phi}_c^{} - \vop_1' \tilde{\phi}_h^{} + \uop_1' \tilde{\phi}_e^{} \;=\; 0\\
\\
\ds \tilde{\ell}_2^{\prime} \; \stackrel{def}{=}\; \tilde{\phi}_h^{} - \vop_2' \tilde{\phi}_b^{} + \uop_2' \tilde{\phi}_d^{} \;=\; 0\\
\\
\ds \tilde{\ell}_3^{\prime} \; \stackrel{def}{=}\; \tilde{\phi}_c^{} - \vop_3' \tilde{\phi}_g^{} + \uop_3' \tilde{\phi}_h^{} \;=\; 0
\end{array}
\right.
\end{equation}
Linear problems (\ref{LP1}) and (\ref{LP2}) are compatible, so that the map $\mathfrak{w}_j^{}\to\mathfrak{w}_j'$ becomes unique:
\begin{equation}\label{Rmap1}
\begin{array}{lll}
\ds \uop_1'\;=\;\uop_2^{}+\uop_1^{}\vop_2^{}\vop_3^{-1}\;, & \ds \uop_2'\;=\;\uop_3^{} (\vop_1^{}+\uop_3^{})^{-1}\uop_1^{}\;, & \ds \uop_3' \;=\; \uop_1^{-1}(\vop_1^{}+\uop_3^{}) \uop_2^{}\;,\\
\\
\ds \vop_1'\;=\;\vop_3^{-1}(\vop_1^{}+\uop_3^{}) \vop_2^{}\;, & \ds \vop_2'\;=\;\vop_1^{}(\vop_1^{}+\uop_3^{})^{-1}\vop_3^{}\;, & \ds \vop_3'\;=\;\vop_2^{}+\uop_1^{-1}\uop_2^{}\vop_3^{}\;.
\end{array}
\end{equation}
This set of formulas defines the map $\mathfrak{R}_{123}$ on a space of functions of $\uop_j,\vop_j$,
\begin{equation}
(\mathfrak{R}_{123}^{}\circ f) (\mathfrak{w}_1^{},\mathfrak{w}_2^{},\mathfrak{w}_3^{})\;=\;
f(\mathfrak{w}_1',\mathfrak{w}_2',\mathfrak{w}_3')\;.
\end{equation}
\begin{statement} Properties of $\mathfrak{R}$ are:
\begin{itemize}
\item The map $\mathfrak{R}$ is the automorphism of $\mathfrak{w}_1\otimes \mathfrak{w}_2\otimes \mathfrak{w}_3$ (proof by direct verification).
\item The map $\mathfrak{R}_{123}$ satisfies the functional Tetrahedron equation (proof by direct verification).
\item In all cases of the self-dual representation of Weyl algebra (e.g. the cyclic $\mathbb{Z}_N$ representation, modular double, $\mathbb{Z}\otimes \mathbb{T}$, etc.), when the quantum dilogarithm is well defined, the quantum operator $R_{123}$, such that
\begin{equation}
(\mathfrak{R}_{123}^{}\circ f) \;=\; R_{123}^{} \cdot f \cdot R_{123}^{-1}\;,
\end{equation}
is uniquely defined. Operator $R_{123}$ has the structure (\ref{R}) and satisfy the quantum Tetrahedron equation \cite{KSprep}.
\end{itemize}
\end{statement}

\section{Evolution map and the integrals of motion}

The evolution map appears as an extension of the ``Yang-Baxter-type map'' (\ref{LYBE}) to the whole kagome lattice:
\begin{equation}\label{EV}
\begin{tikzpicture}
\draw [-latex, thin] (-3,0) -- (1,0);
\draw [-latex, thin] (-3,-1) -- (1,3);
\draw [-latex, thin] (0,-1) -- (0,3);
\node [above] at (-0.55,0.27) {$a_{i,j}$};
\node [above] at (-2,1.5) {$c_{i,j}$};
\node [above] at (-3.3,-0.8) {$b_{i,j-1}$};
\node [above] at (0.7,3.0) {$b_{i-1,j}$};
\node [above] at (1,1) {$c_{i,j+1}$};
\node [above] at (0.8,-1) {$b_{i,j}$};
\node [above] at (-1.3,-1.2) {$c_{i+1,j}$};
\draw [fill] (0,2) circle [radius=0.05];
\draw [fill] (0,0) circle [radius=0.05];
\draw [fill] (-2,0) circle [radius=0.05];
\node [left] at (-0.05,2.05) {$\mathfrak{w}_{1,i,j}$};
\node [above] at (-2.3,0.1) {$\mathfrak{w}_{3,i,j}$};
\node [right] at (0.,0.25) {$\mathfrak{w}_{2,i,j}$};
\node [right] at (2.5,1) {$\ds \stackrel{\mathcal{U}}{\to}$};
\end{tikzpicture}
\qquad
\begin{tikzpicture}
\draw [-latex, thin] (-1,0) -- (3,0);
\draw [-latex, thin] (-1,-3) -- (3,1);
\draw [-latex, thin] (0,-3) -- (0,1);
\node [above] at (0.6,-1) {$b_{i,\j}'$};
\node [above] at (-0.7,0.4) {$a'_{i,j}$};
\node [above] at (-1.1,-1.4) {$c'_{i+1,j}$};
\node [above] at (1.5,0.8) {$c'_{i,j+1}$};
\node [above] at (3.5,0.2) {$a'_{i,j+1}$};
\node [above] at (2.5,-2) {$c'_{i+1,j+1}$};
\node [above] at (-0.6,-3.7) {$a'_{i+1,j}$};
\draw [fill] (0,-2) circle [radius=0.05];
\draw [fill] (0,0) circle [radius=0.05];
\draw [fill] (2,0) circle [radius=0.05];
\node [right] at (0.,-2.1) {$\mathfrak{w}_{1,i+1,j}'$};
\node [below] at (2.5,-0.) {$\mathfrak{w}_{3,i,j+1}'$};
\node [right] at (0.,0.25) {$\mathfrak{w}_{2,i,j}'$};
\end{tikzpicture}
\end{equation}
For the whole kagome lattice, $c_{i,j}$ label the hexagonal cells, while $a_{i,j}$ and $b_{i,j}$ label two types of triangular cells.
Notations for the cells here preserve the structure of the kagome lattice, however some cells must coincide according to (\ref{LYBE}):
\begin{equation}
a_{i,j}'\;=\;c_{i,j}^{}\;,\quad c_{i,j}'\;=\;b_{i-1,j-1}^{}\;,
\end{equation}
while the cells $a_{i,j}$ on the left disappear and cells $b_{i,j}'$ on the right appear.

Explicit form of the evolution map follows from (\ref{Rmap1}):
\begin{equation}\label{Umap}
\begin{array}{ll}
\ds \uop_{1,i+1,j}'\;=\;\left(\uop_2^{}+\uop_1^{}\vop_2^{}\vop_3^{-1}\right)_{i,j}\;, & \ds \vop_{1,i+1,j}'\;=\;\left(\vop_3^{-1}(\vop_1^{}+\uop_3^{}) \vop_2^{}\right)_{i,j}\;,\\
\\
\ds \uop_{2,i,j}'\;=\;\left(\uop_3^{} (\vop_1^{}+\uop_3^{})^{-1}\uop_1^{}\right)_{i,j}\;, & \ds \vop_{2,i,j}'\;=\;\left(\vop_1^{}(\vop_1^{}+\uop_3^{})^{-1}\vop_3^{}\right)_{i,j}\;,\\
\\
\ds \uop_{3,i,j+1}' \;=\; \left(\uop_1^{-1}(\vop_1^{}+\uop_3^{}) \uop_2^{}\right)_{i,j}\;, & \ds \vop_{3,i,j+1}'\;=\;\left(\vop_2^{}+\uop_1^{-1}\uop_2^{}\vop_3^{}\right)_{i,j}\;.
\end{array}
\end{equation}
The evolution map $\ds (\mathcal{U}\circ f)(\mathfrak{w}_{\alpha,i,j}^{}) = f(\mathfrak{w}'_{\alpha,i,j})$, being realised for a particular representation of the Weyl algebra,
$(\mathcal{U}\circ f) \;=\; U \cdot f \cdot U^{-1}$, is given by (\ref{U}).

Generating functions for the integrals of motion preserved by the evolution map are quantum spectral polynomials. 

To obtain the first spectral polynomial, consider the whole set of auxiliary linear relations for the whole kagome lattice. The type (\ref{LP1}) gives
\begin{equation}\label{LP-1}
\left\{ \begin{array}{l}
\ds \ell_{1,i,j}\;=\; \phi_{c_{i,j}}-\vop_{1,ij} \phi_{a_{i,j}}+\uop_{1,ij} \phi_{b_{i-1,j}}\;,\\
\\
\ds \ell_{2,ij}\;=\; \phi_{a_{i,j}} - \vop_{2,ij} \phi_{c_{i+1,j}} + \vop_{2,ij} \phi_{c_{i,j+1}}\;,\\
\\
\ds \ell_{3,ij} \;=\; \phi_{c_{i,j}} - \vop_{3,ij} \phi_{b_{i,j-1}} + \uop_{3,ij} \phi_{a_{i,j}}\;.
\end{array}
\right.
\end{equation}
Consider the periodic boundary conditions and introduce the quasi–periods,
\begin{equation}
\phi_{c_{i+N,j}}\;=\;\lambda \phi_{c_{i,j}}\;,\quad \phi_{c_{i,j+N}}\;=\;\mu\phi_{c_{i,j}}\;,\quad \textrm{etc.}
\end{equation}
The system of linear relations (\ref{LP-1}) can be written in the matrix form,
\begin{equation}
\ell_{A} \;=\; \sum_{B}L_{A,B}^{(1)} \phi_{B}
\end{equation}
where $A$ takes $3N^2$ values $A\in \{(1,ij), (2,ij), (3,ij)\}$, and $B$ takes $3N^2$ values
$B\in \{a_{ij},b_{ij},c_{ij}\}$.

Define now
\begin{equation}\label{det}
J^{(1)}(\lambda,\mu) \;=\; \left( \prod_{i,j} \uop_{1,ij}^{-1}\right) \cdot \det L^{(1)}
\end{equation}
Operator $J^{(1)}(\lambda,\mu)$ is a polynomial of $\lambda,\mu$ with the following Newton polygon (this is a particular case of $N=3$):
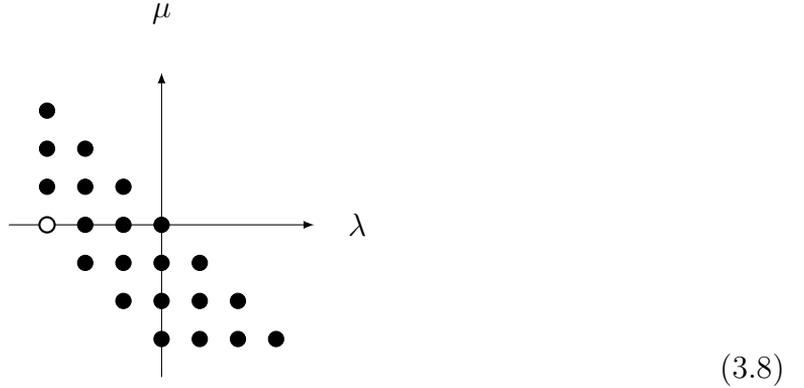
\begin{equation}\label{Newton}
\begin{tikzpicture}
\draw [-latex, thin] (-2,0) -- (2,0);
\node [right] at (2.3,0) {$\lambda$};
\draw [-latex, thin] (0,-2) -- (0,2);
\node [above] at (0,2.5) {$\mu$};
\draw [fill] (0,0) circle [radius=0.1];
\draw [fill] (-0.5,0) circle [radius=0.1];
\draw [fill] (-1,0) circle [radius=0.1];
\draw [fill, white] (-1.5,0) circle [radius=0.1];
\draw [thick] (-1.5,0) circle [radius=0.1];
\draw [fill] (-1.5,0.5) circle [radius=0.1];
\draw [fill] (-1.5,1) circle [radius=0.1];
\draw [fill] (-1.5,1.5) circle [radius=0.1];
\draw [fill] (-1,-0.5) circle [radius=0.1];
\draw [fill] (-1,0.5) circle [radius=0.1];
\draw [fill] (-1,1) circle [radius=0.1];
\draw [fill] (-0.5,-1) circle [radius=0.1];
\draw [fill] (-0.5,-0.5) circle [radius=0.1];
\draw [fill] (-0.5,0.5) circle [radius=0.1];
\draw [fill] (-0,-0.5) circle [radius=0.1];
\draw [fill] (-0,-1) circle [radius=0.1];
\draw [fill] (-0,-1.5) circle [radius=0.1];
\draw [fill] (0.5,-0.5) circle [radius=0.1];
\draw [fill] (0.5,-1) circle [radius=0.1];
\draw [fill] (0.5,-1.5) circle [radius=0.1];
\draw [fill] (1,-1) circle [radius=0.1];
\draw [fill] (1,-1.5) circle [radius=0.1];
\draw [fill] (1.5,-1.5) circle [radius=0.1];
\end{tikzpicture}
\end{equation}
so that
\begin{equation}
J^{(1)}(\lambda,\mu)\;=\;\sum_{a,b} \lambda^a\mu^b J^{(1)}_{a,b}
\end{equation}
The extra multiplier in (\ref{det}) makes
\begin{equation}
J^{(1)}_{-N,0}\;=\;1\;,
\end{equation}
this coefficient corresponds to the empty dot in (\ref{Newton}).

In the similar way one can consider the linear problem (\ref{LP2}),
\begin{equation}\label{LP-2}
\left\{ \begin{array}{l}
\ds \tilde{\ell}_{1,i,j}\;=\; \tilde{\phi}_{c_{i,j+1}}-\vop_{1,ij} \tilde{\phi}_{b_{i-1,j}}+\uop_{1,ij} \tilde{\phi}_{a_{i,j}}\;,\\
\\
\ds \tilde{\ell}_{2,ij}\;=\; \tilde{\phi}_{b_{i,j}} - \vop_{2,ij} \tilde{\phi}_{c_{i,j+1}} + \vop_{2,ij} \tilde{\phi}_{c_{i+1,j}}\;,\\
\\
\ds \tilde{\ell}_{3,ij} \;=\; \tilde{\phi}_{c_{i+1,j}} - \vop_{3,ij} \tilde{\phi}_{a_{i,j}} + \uop_{3,ij} \tilde{\phi}_{b_{i,j-1}}\;,
\end{array}
\right.
\end{equation}
but now with the opposite quasimomenta,
\begin{equation}
\tilde{\phi}_{c_{i+N,j}}\;=\;\lambda^{-1} \tilde{\phi}_{c_{i,j}}\;,\quad \tilde{\phi}_{c_{i,j+N}}\;=\;\mu^{-1}\tilde{\phi}_{c_{i,j}}\;,\quad \textrm{etc.}
\end{equation}
In the same way we can define its determinant,
\begin{equation}
\tilde{\ell}_{A}\;=\;\sum_{B} L^{(2)}_{A,B} \tilde{\phi}_{B}\;,\quad J^{(2)}(\lambda,\mu) \;=\; \left(\prod_{i,j} \uop_{1,ij}^{-1}\right) \cdot \det L^{(2)}\;.
\end{equation}
Polynomial $J^{(2)}(\lambda,\mu)$ has the same Newton polygon (\ref{Newton}).
\begin{statement}
Polynomials $J^{(k)}(\lambda,\mu)$, $k=1,2$, have the following properties:
\begin{itemize}
\item They generate the integrals of motion, $\mathcal{U}\circ J^{(k)}(\lambda,\mu) \;=\; J^{(k)}(\lambda,\mu)$.
\item Polynomials $J^{(1)}(\lambda,\mu)$ and $J^{(2)}(\lambda,\mu)$ have identical perimeters.
\item Elements $J^{(1,2)}_{a,b}$ and from the inside of the Newton polygon are algebraically independent.
\item Elements $J^{(1,2)}_{a,b}$ satisfy the algebra
\begin{equation}\label{algebra}
\ds J^{(k)}_{a,b} J^{(k')}_{a',b'}\;=\;q^{2(a'+N)b-2(a+N)b'} J^{(k')}_{a',b'} J^{(k)}_{a,b}\;.
\end{equation}
\item Thus, the set $J^{(1,2)}_{a,b}$ gives $3N^2+1$ invariants with $3N^2$ commutative independent invariants.
\end{itemize}
Thus, the Liouville integrability of the evolution is established.
\end{statement}
\noindent
\textbf{Comments on the Statement.} The general approach to the quantum curve method, including the proof of (\ref{algebra}), can be found in \cite{S2,S3}. However, in this paper two different quantum curves appear, so that all items from the statement above, especially the statements about completeness, have been verified straightforwardly for $N \leq 4$.

Common perimeters of $J^{(1,2)}(\lambda,\mu)$ worth mentioning. The most left side of (\ref{Newton}) is given by
\begin{equation}
J^{(1,2)}(\lambda,\mu) \;=\; \lambda^{-N} \prod_i \left( 1 - \mu \prod_j \uop_{2,ij}\uop_{3,ij} \right) + \lambda^{-N+1}\cdots\;,
\end{equation}
the bottom side is given by
\begin{equation}
J^{(1,2)}(\lambda,\mu)\;=\;\mu^{-N} \left(\prod_{i,j} \uop_{1,ij}^{-1}\vop_{3,ij}^{}\right) \,
\prod_{j} \left( 1 - \lambda \prod_i \vop_{1,ij} \vop_{2,ij} \right) + \mu^{-N+1}\cdots\;,
\end{equation}
and the left-botton side is given by
\begin{equation}
J^{(1,2)}(\lambda,\mu) \;=\; \lambda^{-N}\prod_{k} \left( 1+ \frac{\lambda}{\mu} \prod_{i+j=k} \uop_{1,ij}^{-1}\vop_{3,ij}^{}\right)+\cdots
\end{equation}
One can extract very simple invariants,
\begin{equation}
\boldsymbol{U}_i\;=\;\prod_{j} \uop_{2,ij}\uop_{3,ij}\;,\quad 
\boldsymbol{V}_j\;=\;\prod_{i} \vop_{1,ij}\vop_{2,ij}\;,
\end{equation}
so that
\begin{equation}
I^{(k)}_{a,b}\;=\;\boldsymbol{U}_0^{-b} \boldsymbol{V}_0^{-a-N} J^{(k)}_{a,b}\;,
\end{equation}
is the set of integrals of motion in involution. The non-commutative pair $\boldsymbol{U},\boldsymbol{V}$ has the meaning of a mass center of the system.

\section{Classical model}

It is interesting to demonstrate the existence of two independent spectral curves on the classical level.

To do it, we rewrite firstly the evolution (\ref{Umap}) in the form of equations of motion on the cubic lattice.

Let $\pt=(n_1,n_2,n_3)$ stand for a vertex of the cubic lattice. Let 
\begin{equation}\label{3dcoord}
\one=(1,0,0)\;,\quad \two=(0,1,0)\;,\quad \three=(0,0,1)
\end{equation}
be unit vectors. Then the local map (\ref{Rmap1}) can be written as 
\begin{equation}\label{eom}
\begin{array}{ll}
\ds u_{1,\pt+\one}\;=\;\frac{u_{2,\pt} v_{3,\pt}+u_{1,\pt} v_{2,\pt}}{v_{3,\pt}}\;, & 
\ds v_{1,\pt+\one}\;=\;\frac{v_{2,\pt}(v_{1,\pt}+u_{3,\pt})}{v_{3,\pt}}\;, \\
\\
\ds u_{2,\pt+\two}\;=\; \frac{u_{1,\pt}u_{3,\pt}}{v_{1,\pt}+u_{3,\pt}}\;, & 
\ds v_{2,\pt+\two}\;=\;\frac{v_{1,\pt}v_{3,\pt}}{v_{1,\pt}+u_{3,\pt}}\;, \\
\\
\ds u_{3,\pt+\three} \;=\; \frac{u_{2,\pt}(v_{1,\pt}+u_{3,\pt})}{u_{1,\pt}}\;, & 
\ds v_{3,\pt+\three}\;=\;\frac{u_{2\pt}v_{3,\pt}+u_{1,\pt}v_{2,\pt}}{u_{1,\pt}}\;.
\end{array}
\end{equation}
The evolution map (\ref{Umap}) is the map $t\to t+1$ of the space-like two dimensional surface
\begin{equation}\label{ijt}
\pt\;=\;(i,t-i-j,j)\;,\quad t-\textrm{fixed.}
\end{equation}
The Legendre transform bringing the Hamiltonian equations of motion (\ref{eom}) to the corresponding Lagrangian equations is not quite evident. It is
\begin{equation}
\begin{array}{ll}
\ds u_{1,\pt}\;=\; 
\frac{\tau^{(1)}_{\pt}}{\tau^{(1)}_{\pt+\three}}
\frac{\tau^{(2)}_{\pt+\two+\three}}{\tau^{(2)}_{\pt+\two}} 
\;, & 
\ds 
v_{1,\pt}\;=\; 
\frac{\tau^{(1)}_{\pt+\two+\three}}{\tau^{(1)}_{\pt+\three}}
\frac{\tau^{(2)}_{\pt}}{\tau^{(2)}_{\pt+\two}}
\;,\\
\\
\ds 
u_{2,\pt}\;=\;  
\frac{\tau^{(1)}_{\pt+\one}}{\tau^{(1)}_{\pt+\one+\three}}
\frac{\tau^{(2)}_{\pt+\three}}{\tau^{(2)}_{\pt}} 
\;, & 
\ds 
v_{2,\pt}\;=\;  
\frac{\tau^{(1)}_{\pt+\three}}{\tau^{(1)}_{\pt+\one+\three}}
\frac{\tau^{(2)}_{\pt+\one}}{\tau^{(2)}_{\pt}}
\;,\\
\\
\ds 
u_{3,\pt}\;=\; 
\frac{\tau^{(1)}_{\pt+\one+\two}}{\tau^{(1)}_{\pt+\one}}
\frac{\tau^{(2)}_{\pt}}{\tau^{(2)}_{\pt+\two}}
\;, & 
\ds 
v_{3,\pt} \;=\; 
\frac{\tau^{(1)}_{\pt}}{\tau^{(1)}_{\pt+\one}}
\frac{\tau^{(2)}_{\pt+\one+\two}}{\tau^{(2)}_{\pt+\two}}
\;.
\end{array}
\end{equation}
As the result, the Lagrangian equations of motion decouple into two independent Hirota equations:
\begin{equation}
\begin{array}{l}
\ds \tau^{(1)}_{\pt} \tau^{(1)}_{\pt+\one+\two+\three}\;=\;  \tau^{(1)}_{\pt+\one} \tau^{(1)}_{\pt+\two+\three} +\tau^{(1)}_{\pt+\three} \tau^{(1)}_{\pt+\one+\two}\;,\\
\\
\ds \tau^{(2)}_{\pt} \tau^{(2)}_{\pt+\one+\two+\three}\;=\; \tau^{(2)}_{\pt+\one} \tau^{(2)}_{\pt+\two+\three} + \tau^{(2)}_{\pt+\three} \tau^{(2)}_{\pt+\one+\two}\;.
\end{array}
\end{equation}
The algebraic curve $J^{(k)}(\lambda,\mu)=0$ becomes the spectral curve for $\tau^{(k)}$-equation, $k=1,2$.

\vspace{1cm}

\noindent
\textbf{Acknowledgement.} I would like to thank Vladimir Bazhanov and Vladimir Mangazeev for valuable discussions.
Also I acknowledge the support of the Australian Research Council grant 
DP190103144.

\appendix

\appendixpage

There is the term ``Pentagon equation'' in the title of this paper, however if fact we did not use it. In this Appendix we give some alternative way of derivation of the Tetrahedral map (\ref{Rmap1}) using different linear problem. This approach is more geometric, but it does not allowed one to derive the spectral determinants. In addition, we collect here some by-product observations which may be interesting for the Reader.

\section{Pentagon Map}

We start with the Pentagon map and underlying auxiliary linear problem \cite{DS1}.

Let $\uop,\vop$ be the simple Weyl algebra,
\begin{equation}
\uop\,\vop\;=\;q^2\,\vop\,\uop\;.
\end{equation}
The quantum Pentagon map is defined over the space of functions on the direct product of two Weyl algebras,
\begin{equation}
\biggl( \mathcal{S}_{12}^{} \circ \varphi\biggr) (\uop_1^{},\vop_1^{},\uop_2^{},\vop_2^{})\;=\;
\varphi(\uop_1',\vop_1',\uop_2',\vop_2')\qquad \forall \varphi
\end{equation}
where
\begin{equation}\label{Smap}
\begin{array}{ll}
\ds \uop_1'\;=\;\vop_2^{-1}\uop_1^{}\;,& \ds \vop_1'\;=\;\vop_2^{-1}(\vop_1^{}+\uop_2^{})\;,\\
\\
\ds \uop_2'\;=\;\uop_2^{}(\vop_1^{}+\uop_2^{})^{-1}\uop_1^{}\;, & \ds \vop_2'\;=\;\vop_1^{}(\vop_1^{}+\uop_2^{})^{-1}\vop_2^{}\;.
\end{array}
\end{equation}
The map $\mathcal{S}_{12}$ is the canonical one, i.e. it can be seen as a conjugation,
\begin{equation}
\biggl(\mathcal{S}_{12}\circ\varphi\biggr)\;=\;S_{12}^{}\,\varphi\, S_{12}^{-1}\;.
\end{equation}
Construction of the kernel of $S_{12}$ is the subject of the choice of the representation of the Weyl algebra and quantum dilogarithms. However, this is not a subject of this paper. We focus here on the map (\ref{Smap}) only.

The map (\ref{Smap}) can be derived as the consistency condition of the following ``linear'' problem. Consider triangle $ABC$,
\begin{equation}\label{triangle}
\begin{tikzpicture}[scale=1.2]
\draw [ultra thick] (0,0) -- (1,1) -- (1,-1) --  (0,0);
\draw [fill] (0,0) circle [radius=0.05];
\draw [fill] (1,1) circle [radius=0.05];
\draw [fill] (1,-1) circle [radius=0.05];
\node [below] at (1,-1) {$A$};
\node [left] at (0,0) {$B$};
\node [above] at (1,1) {$C$};
\draw (1,-0.7) arc [start angle=90, end angle=135, radius=0.3];
\node [right] at (2,0) {$\Leftrightarrow\quad \varphi_A \,-\, \vop\,\varphi_B \,+\, \uop \, \varphi_C\;=\;0\;.$};
\end{tikzpicture}
\end{equation}
Let a formal linear variable $\phi_A$ is associated with the vertex $A$, etc. For the whole triangle we associate the linear equation shown in the left part of (\ref{triangle}). Note the orientation of the triangle, vertex $A$ is selected, the vertices $ABC$ are oriented clockwise. 

Consider now quadrilateral $ABCD$. It can be divided into two triangles in two ways:
\begin{equation}\label{quad}
\begin{tikzpicture}[scale=1.2]
\draw [ultra thick] (0,0) -- (1,1) -- (2,0) -- (1,-1) -- (0,0);
\draw [ultra thick] (0,0) -- (2,0);
\draw [fill] (0,0) circle [radius=0.05];
\draw [fill] (1,1) circle [radius=0.05];
\draw [fill] (2,0) circle [radius=0.05];
\draw [fill] (1,-1) circle [radius=0.05];
\node [below] at (1,-1) {$A$};
\node [left] at (0,0) {$B$};
\node [above] at (1,1) {$C$};
\node [right] at (2,0) {$D$};
\node [right] at (0.8,0.4) {$1$};
\node [right] at (0.8,-0.4) {$2$};
\draw (1.7,0) arc [start angle=180, end angle=135, radius=0.3];
\draw (1.7,0) arc [start angle=180, end angle=225, radius=0.3];
\end{tikzpicture}
\qquad
\begin{tikzpicture}[scale=1.2]
\node [left] at (-1.2,0) {$\sim$};
\draw [ultra thick] (0,0) -- (1,1) -- (2,0) -- (1,-1) -- (0,0);
\draw [ultra thick] (1,1) -- (1,-1);
\draw [fill] (0,0) circle [radius=0.05];
\draw [fill] (1,1) circle [radius=0.05];
\draw [fill] (2,0) circle [radius=0.05];
\draw [fill] (1,-1) circle [radius=0.05];
\node [below] at (1,-1) {$A$};
\node [left] at (0,0) {$B$};
\node [above] at (1,1) {$C$};
\node [right] at (2,0) {$D$};
\node [right] at (0.4,0) {$1$};
\node [right] at (1.2,0) {$2$};
\draw (1,-0.7) arc [start angle=90, end angle=135, radius=0.3];
\draw (1.7,0) arc [start angle=180, end angle=135, radius=0.3];
\draw (1.7,0) arc [start angle=180, end angle=225, radius=0.3];
\end{tikzpicture}
\end{equation}
The left hand side of (\ref{quad}) implies two linear equations
\begin{equation}\label{ql}
\varphi_D - \vop_1^{}\varphi_B + \uop_1^{} \varphi_C\;=\;0\;,\quad
\varphi_D - \vop_2^{}\varphi_A + \uop_2^{}\varphi_B\;=\;0\;.
\end{equation}
For the right hand side of (\ref{quad}) we write the similar equations,
\begin{equation}\label{qr}
\varphi_A - \vop_1'\varphi_B + \uop_1'\varphi_C\;=\;0\;.\quad
\varphi_D - \vop_2'\varphi_A + \uop_2'\varphi_C\;=\;0\;,
\end{equation}
The map (\ref{Smap}) appear as the equivalence requirement of (\ref{ql}) and (\ref{qr}). For instance, excluding $\varphi_D$ from (\ref{ql}), one obtains
\begin{equation}
\varphi_A-\vop_2^{-1}(\vop_1^{}+\uop_2^{})\varphi_B+\vop_2^{-1}\uop_1^{}\varphi_C\;=\;0\;.
\end{equation} 
This equation could be identically the left equation of (\ref{qr}), what gives the expressions (\ref{Smap}) for $\vop_1'$ and $\uop_1'$. Similarly, excluding $\varphi_B$ from (\ref{ql}), one obtains expressions for $\vop_2',\uop_2'$.

Note that the map (\ref{Smap}) is constructed as the map of a generic non-commutative algebra of observables. By construction, it satisfies the algebraic Pentagon relation (\ref{PE}) as well as the ten-term relation (\ref{10}) since all these relations can be formulated as uniquely defined maps for corresponding auxiliary linear problems.

In particular, all maps in this paper are the canonical maps with respect to the local Weyl algebra:
\begin{equation}\label{Wloc}
\uop_j^{}\,\vop_j^{}\;=\;q^2\,\vop_j^{}\uop_j^{}\quad \Rightarrow\quad \uop_j'\,\vop_j'\;=\;q^2\,\vop_j'\,\uop_j'\;,
\end{equation}
while the elements with different $j$ commute. The classical limit of the local Weyl algebra implies the conservation of the canonical form
\begin{equation}\label{Ploc}
\sum_j d\log u_j^{}\wedge d\log v_j^{} \;=\;
\sum_j d\log u_j'\wedge d\log v_j' \;.
\end{equation}

There is a temptation to give some geometric interpretation to the map $S_{12}$ and to Pentagon and ten-term relations using the idea that the quadrilateral $ABCD$ for the map $S_{12}$ is a tetrahedron, the Pentagon relation (\ref{PE}) corresponds to a dissection of a four-simplex into five tetrahedra, and so on. However, the attempt to give geometric interpretation failed. In what follows, we will understand the linear problems only in the combinatorial way. For the shortness, we will use notation
\begin{equation}\label{triple}
(ABC)_j^{\#}\quad \Leftrightarrow \quad \phi_A-\vop_j^{\#}\phi_B+\uop_j^{\#}\phi_C\;=\;0\;.
\end{equation}

\section{Tetrahedral  Map}

The Tetrahedral map (\ref{Rmap1}) appears as the equivalence of
\begin{equation}\label{Rlin}
\left\{\begin{array}{l}
\ds (DBC)_1^{}\\
\ds (EAC)_2^{}\\
\ds (DAB)_3^{}
\end{array}
\right. \quad \sim\quad
\left\{\begin{array}{l}
\ds (EBC)_1'\\
\ds (DAC)_2'\\
\ds (EAB)_3'
\end{array}
\right.
\end{equation}
Precise answer for $\mathcal{R}_{123}^{}\;:\;\uop_j^{},\vop_j^{}\;\to\;\uop_j',\vop_j'$ is
\begin{equation}\label{Rmap2}
\begin{array}{lll}
\ds \uop_1'\;=\;\uop_2^{}+\vop_2^{}\vop_3^{-1}\uop_1^{}\;,& 
\ds \uop_2'\;=\;\uop_3^{}(\vop_1^{}+\uop_3)^{-1}\uop_1^{}\;,&
\ds \uop_3'\;=\;\uop_2^{}\uop_1^{-1}(\vop_1^{}+\uop_3^{})\;,\\
\\
\ds \vop_1'\;=\;\vop_2^{}\vop_3^{-1}(\vop_1^{}+\uop_3^{})\;,&
\ds \vop_2'\;=\;\vop_1^{}(\vop_1^{}+\uop_3^{})^{-1}\vop_3^{}\;,&
\ds \vop_3'\;=\;\vop_2^{}+\uop_2^{}\uop_1^{-1}\vop_3^{}\;.
\end{array}
\end{equation}
The reason why we repeat (\ref{Rmap1}) here is that the map (\ref{Rmap2}) satisfies the functional Tetrahedron equation in the terms of free algebra of $\uop_i,\vop_j$, $j=1,\cdots,6$, not necessarily the ultra-local algebra.

The linear problems (\ref{Rlin}) has the following combinatorics:
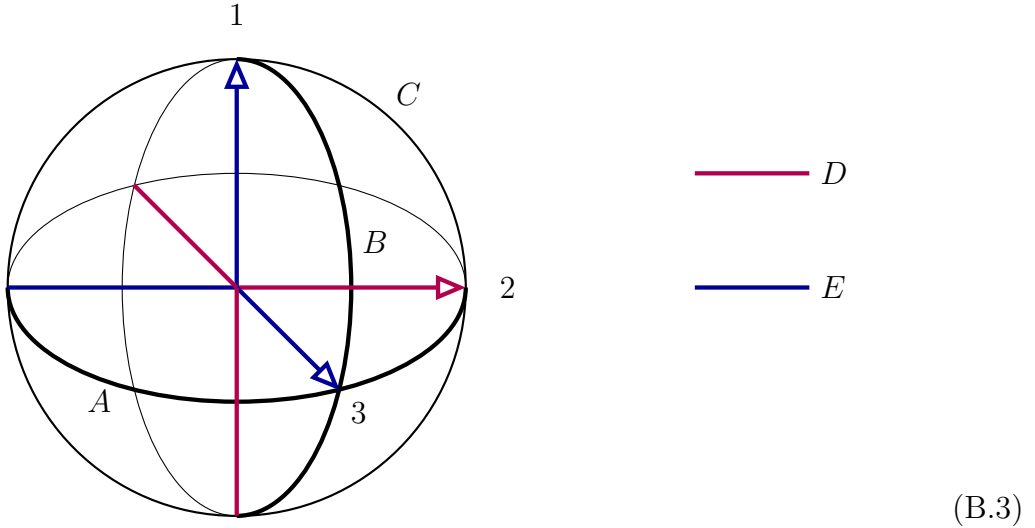
\begin{equation}\label{picture}
\begin{tikzpicture}[scale=3]
\draw [thick] (0,0) circle (1);
\draw [ultra thick] (0,-1) arc (-90:90:0.5 and 1);
\draw [thin] (0,1) arc (90:270:0.5 and 1);
\draw [ultra thick] (-1,0) arc (-180:0:1 and 0.5);
\draw [thin] (1,0) arc (0:180:1 and 0.5);
\draw [-open triangle 45, ultra thick, \GR] (0,0) -- (1,0);
\draw [-open triangle 45, ultra thick, \PAC] (0,0) -- (0,1);
\draw [-open triangle 45, ultra thick, \PAC] (0,0) -- (0.45,-0.45);
\draw [ultra thick, \PAC] (0,0) -- (-1,0);
\draw [ultra thick, \GR] (0,0) -- (0,-1);
\draw [ultra thick, \GR] (0,0) -- (-0.45,0.45);
\node [above] at (0,1.1) {$1$};
\node [right] at (1.1,0) {$2$};
\node [right] at (0.45,-0.55) {$3$};
\node [above] at (0.75,0.75) {$C$};
\node [right] at (0.5,0.20) {$B$};
\node [below] at (-0.6,-0.4) {$A$};
\draw [ultra thick, \GR] (2,0.5) -- (2.5,0.5);
\node [right] at (2.5,0.5) {$D$};
\draw [ultra thick, \PAC] (2,0) -- (2.5,0);
\node [right] at (2.5,0) {$E$};
\end{tikzpicture}
\end{equation} 
Here $1,2,3$ are the  indices of $\uop,\vop$ in the linear equations. Incoming edges correspond to $\uop_j,\vop_j$, while the outgoing edges correspond to $\uop_j',\vop_j'$. Edge $1$ is the intersection of planes $B$ and $C$, edge $2$ is the intersection of planes $A$ and $C$, and edge $3$ is the intersection of planes $A$ and $B$. The edges have two ``colours'', $D$ and $E$, as it is shown in (\ref{picture}).

These combinatorial rules allows one to extend the linear problems to the whole cubic lattice.

\section{Evolution map}

The evolution map is given by (\ref{Umap}). Here we repeat it with the time variable shown explicitly:
\begin{equation}\label{Umap2}
\begin{array}{ll}
\ds \uop_{1,(i+1,j)}^{(t+1)}\;=\;\left(\uop_2^{}+\vop_2^{}\vop_3^{-1}\uop_1^{}\right)_{(i,j)}^{(t)}\;,& 
\ds \vop_{1,(i+1,j)}^{(t+1)}\;=\;\left(\vop_2^{}\vop_3^{-1}(\vop_1^{}+\uop_3^{})\right)_{(i,j)}^{(t)}\;,\\
\\
\ds \uop_{2,(i,j)}^{(t+1)}\;=\;\left(\uop_3^{}(\vop_1^{}+\uop_3)^{-1}\uop_1^{}\right)_{(i,j)}^{(t)}\;,&
\ds \vop_{2,(i,j)}^{(t+1)}\;=\;\left(\vop_1^{}(\vop_1^{}+\uop_3^{})^{-1}\vop_3^{}\right)_{(i,j)}^{(t)}\;,\\
\\
\ds \uop_{3,(i,j+1)}^{(t+1)}\;=\;\left(\uop_2^{}\uop_1^{-1}(\vop_1^{}+\uop_3^{})\right)_{(i,j)}^{(t)}\;,&
\ds \vop_{3,(i,j+1)}^{(t+1)}\;=\;\left(\vop_2^{}+\uop_2^{}\uop_1^{-1}\vop_3^{}\right)_{(i,j)}^{(t)}\;.
\end{array}
\end{equation}
\\

The set of all auxiliary linear equations can be described by the following set of the triples,
\begin{equation}\label{system}
(D_{i+k+j},B_k,C_j)_{1,(i,j)}\;,\quad
(D_{i+k+j+1},A_i,C_j)_{2,(i,j)}\;,\quad
(D_{i+k+j},A_i,B_k)_{3,(i,j)}\;,
\end{equation}
where we used notation of eq. (\ref{triple}), and $t\;=\;i-k+j$. Periodical boundary conditions $i,j\in\mathbb{Z}_N$ require the periodicity of $D_{i+k+j}$. Note that the coordinate $\pt$ (\ref{3dcoord}) is given by $\pt=(i,-k,j)$, see also (\ref{ijt}).

One can see now, the auxiliary liner problem (\ref{system}) at fixed time $t$ is the set of $3N^2$ equations for $4N$ auxiliary linear variables. We did not find any way to relate this system to the integrals of motion (except for $N=1$). In particular, the combinatorics of (\ref{system}) does not allow to interpret it as a set of Fay's identities.

\section{Another example of classical Pentagon map}

We'd like to mention one geometric example of Pentagon and Tetrahedral maps solving the functional Pentagon and Tetrahedron equations. This example is purely classical, it has no Poisson structure and quantum counterpart, but it leads to some classical integrable model on cubic lattice.

We use the framework of Euclidean geometry. Consider the flat quadrilateral (dissected into two triangles in two ways) with the angles shown below:
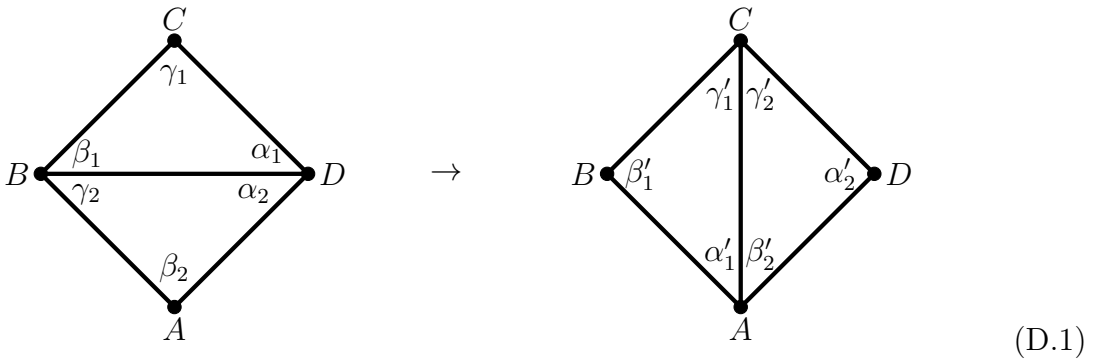
\begin{equation}\label{quad2}
\begin{tikzpicture}[scale=1.75]
\draw [ultra thick] (0,0) -- (1,1) -- (2,0) -- (1,-1) -- (0,0);
\draw [ultra thick] (0,0) -- (2,0);
\draw [fill] (0,0) circle [radius=0.05];
\draw [fill] (1,1) circle [radius=0.05];
\draw [fill] (2,0) circle [radius=0.05];
\draw [fill] (1,-1) circle [radius=0.05];
\node [below] at (1,-1) {$A$};
\node [left] at (0,0) {$B$};
\node [above] at (1,1) {$C$};
\node [right] at (2,0) {$D$};
\node [left] at (1.9,0.15) {$\alpha_1$};
\node [right] at (0.15,0.15) {$\beta_1$};
\node [below] at (1,0.9) {$\gamma_1$};
\node [left] at (1.8,-0.15) {$\alpha_2$};
\node [above] at (1,-0.9) {$\beta_2$};
\node [right] at (0.15,-0.15) {$\gamma_2$};
\end{tikzpicture}
\qquad
\begin{tikzpicture}[scale=1.75]
\node [left] at (-1.,0) {$\to$};
\draw [ultra thick] (0,0) -- (1,1) -- (2,0) -- (1,-1) -- (0,0);
\draw [ultra thick] (1,1) -- (1,-1);
\draw [fill] (0,0) circle [radius=0.05];
\draw [fill] (1,1) circle [radius=0.05];
\draw [fill] (2,0) circle [radius=0.05];
\draw [fill] (1,-1) circle [radius=0.05];
\node [below] at (1,-1) {$A$};
\node [left] at (0,0) {$B$};
\node [above] at (1,1) {$C$};
\node [right] at (2,0) {$D$};
\node [above] at (0.85,-0.8) {$\alpha_1'$};
\node [right] at (0.05,0) {$\beta_1'$};
\node [below] at (0.85,0.8) {$\gamma_1'$};
\node [left] at (1.95,0) {$\alpha_2'$};
\node [above] at (1.15,-0.8) {$\beta_2'$};
\node [below] at (1.15,0.8) {$\gamma_2'$};
\end{tikzpicture}
\end{equation}
Define
\begin{equation}
u_j^{}\;=\;\EXP^{2\ii \beta_j^{}}\;,\quad 
v_j^{}\;=\;\EXP^{2\ii \gamma_j^{}}\;,\quad
u_j'\;=\;\EXP^{2\ii \beta_j'}\;,\quad 
v_j'\;=\;\EXP^{2\ii \gamma_j'}\;.
\end{equation}
The framework of the Euclidean geometry provides
\begin{equation}
u_1'=u_1v_2\;,\quad v_1'=\frac{F}{v_2}\;,\quad u_2'=u_1u_2 F\;,\quad v_2'=\frac{v_1v_2}{F}\;,
\end{equation}
where 
\begin{equation}
F\;=\;-\, \frac{u_1u_2v_1v_2-u_1v_1v_2-u_2v_1v_2+v_1+v_2-1}{u_1u_2v_1v_2-u_1u_2v_1-u_1u_2v_2+u_1+u_2-1}\;.
\end{equation}
It can be verified straightforwardly, $S_{1,2}:u,v\to u',v'$ satisfies the functional Pentagon and ten-term equations, and therefore a solution of the functional Tetrahedron equation can be constructed.

\section{Beyond the Pentagon equation}

This section is an attempt to construct a generalisation of the Pentagon and ten-term relations (\ref{PE},\ref{10}):
\begin{equation}\label{PEg}
(S_c)_{23}^{} (S_b)_{13}^{} (S_a)_{12}^{} \;=\; \sum_{d,e} m_{a,b,c}^{\;d,e} (S_d)_{12}^{} (S_e)_{23}^{}\;,
\end{equation}
\begin{equation}\label{nPEg}
(\overline{S}^{\,a})_{12}^{} (\overline{S}^{\,b})_{13}^{} (\overline{S}^{\,c})_{23}^{} \;=\; \sum_{d,e} \overline{m}^{\,a,b,c}_{\;\,d,e} (\overline{S}^{\,e})_{23}^{} (\overline{S}^{\,d})_{12}^{}\;,
\end{equation}
and
\begin{equation}\label{10g}
\begin{array}{l}
\ds \sum_{b_1,b_2,b_3} \overline{m}^{\, b_1,b_2,b_3}_{\;\,c_1,c_2}
(S_{b_1})_{12}^{} (\overline{S}^{\, a_1})_{13}^{} (S_{b_2})_{14}^{} (\overline{S}^{\,a_2})_{24}^{} (S_{b_3})_{34}^{}\;=\\
\\
\ds \qquad \qquad =\; \sum_{b_1,b_2,b_3}
m_{b_1,b_2,b_3}^{\;a_1,a_2}
(\overline{S}^{\,b_3})_{24}^{} (S_{c_2})_{34}^{} (\overline{S}^{\,b_2})_{14}^{} (S_{c_1})_{12}^{} (\overline{S}^{\,b_1})_{13}^{}\;.
\end{array}
\end{equation}
Due to these (hypothetical) relations
\begin{equation}
R_{123}\;=\;\sum_a (S_{a})_{13}^{} P_{23}^{} (\overline{S}^{\,a})_{13}
\end{equation}
satisfy the Tetrahedron equation, and, moreover,
\begin{equation}
R_{0123}\;=\;\sum_a (S_{a})_{13}^{} P_{01} P_{23}^{} (\overline{S}^{\,a})_{13}
\end{equation}
satisfy the $4$-simplex equation.

The formal ``Pentagonal algebra'' (\ref{PEg}) has a remarkable feature: it implies adjoint representations. Namely, the associativity condition for (\ref{PEg}) reads
\begin{equation}\label{YBEg}
\sum_{b_1,b_2,b_3} m_{a_1,a_2,a_3}^{\; b_1,b_2}
m_{b_1,a_4,a_5}^{\; c_1,b_3}
m_{b_2,b_3,a_6}^{\; c_2,c_3}\;=\;
\sum_{b_1,b_2,b_3}
m_{a_1,b_1,b_2}^{\; c_1,c_2}
m_{a_2,a_4,b_3}^{\; b_1,c_3}
m_{a_3,a_5,a_5}^{\; b_2,b_3}\;.
\end{equation}
The associativity condition for $\overline{m}$ is identical. The adjoint representations are:
\begin{equation}
\begin{array}{ll}
\ds \textrm{(I)}: & \ds m_{a,b,c}^{\;d,e}\;=\; \langle d,e|S_{c} | a,b\rangle \quad \textrm{or} \quad
m_{a,b,c}^{\;d,e}\;=\; \langle b,a|S_{c} | e,d\rangle\;,\\
\\
\ds \textrm{(II)}: & \ds m_{a,b,c}^{\;d,e}\;=\; \langle b,c|S_{a} | d,e\rangle \quad \textrm{or} \quad
m_{a,b,c}^{\;d,e}\;=\; \langle e,d|S_{a} | c,b\rangle\;.
\end{array}
\end{equation} 
So far, we succeeded to find just one example for $m,\overline{m}$ satisfying (\ref{YBEg}) and therefore (\ref{PEg},\ref{nPEg},\ref{10g}) in the adjoint representations:
\begin{equation}
m_{a,b,c}^{\; d,e} \;=\; \overline{m}^{\,a,b,c}_{\;\, d,e} \;=\; 
\delta_{a+b,d}^{}\;  \delta_{b+c,e}^{}\;  q^{2ac}\;,
\end{equation}
where $q\in \mathbb{C}$ for $a,\dots,e\in\mathbb{Z}$ and $q^{2N}\;=\;1$ for $a,\dots,e\in\mathbb{Z}_N$.


\begin{thebibliography}{100}

\bibitem{KS1}
R. M. Kashaev and S. M. Sergeev, ``\emph{On pentagon, ten-term, and tetrahedron relations}'', Comm. Math. Phys. 195 (1998) 309-319


\bibitem{S99}
S. M. Sergeev, ``\emph{Quantum $2 + 1$ evolution model}''. J. Phys. A: Math. Gen. 32 (1999) 5693-5714


\bibitem{S2}
S. M. Sergeev, ``\emph{Coefficient Matrices of a Quantum Discrete Auxiliary Linear Problem}''. Journal of Mathematical Sciences 115(1) (2003) pp 2049-2057 

\bibitem{S3}
S. Sergeev, ``\emph{Quantum curve in q-oscillator model}''. International Journal of Mathematics and Mathematical Sciences Volume 2006, Article ID 92064, Pages 1 -- 31 (2006)

\bibitem{KSprep}
R. M. Kashaev and S. M .Sergeev, \emph{in preparation}

\bibitem{DS1}
A. Doliwa and S. M. Sergeev, ``\emph{The pentagon relation and incidence geometry}''. J. Math. Phys. 55 (2014) 063504


\end{thebibliography}
\end{document}